\documentclass[12pt,epsf,amstex]{article}
\usepackage [dvips]{graphicx}
\usepackage{amsmath}
\usepackage{amssymb}
\usepackage{epsfig}
\usepackage{verbatim}
\usepackage[usenames,dvipsnames,svgnames]{xcolor}

\addtocounter{secnumdepth}{1}
\font\capfont=cmbx12 at 50 pt 
\newbox\capbox \newcount\capl \def\a{A}
\def\docappar{\medbreak\noindent\setbox\capbox\hbox{%
\capfont\a\hskip0.15em}\hangindent=\wd\capbox%
\capl=\ht\capbox\divide\capl by\baselineskip\advance\capl by1%
\hangafter=-\capl%
\hbox{\vbox to8pt{\hbox to0pt{\hss\box\capbox}\vss}}}
\def\cappar{\afterassignment\docappar\noexpand\let\a }

\begin{document}

\newcommand{\ee}{{\rm e}}

\newcommand{\bey}{\boldsymbol{e}_y}
\newcommand{\bv}{\mathbf{v}}

\newcommand{\infy}{\inf_{\rule{0mm}{2.55mm}y}}

\newcommand{\tq}{t_{\rm q}}
\newcommand{\tqm}{t_{\rm q}^*}
\newcommand{\tqs}{t_{\rm qs}}
\newcommand{\tmax}{t_{\rm q}^{\rm max}}
\newcommand{\tauqs}{\tau_{\rm qs}}
\newcommand{\rhotwo}{\rho_2}
\newcommand{\rhoin}{\rho_{\rm in}}
\newcommand{\bE}{\bar{E}}
\newcommand{\xic}{\xi_{\rm min}}
\newcommand{\xiMC}{\xi_{\rm MC}}
\newcommand{\zc}{z_{\rm c}}

\newcommand{\calE}{{\cal E}}
\newcommand{\calH}{{\cal H}}
\newcommand{\calJ}{{\cal J}}
\newcommand{\calK}{{\cal K}}
\newcommand{\calL}{{\cal L}}
\newcommand{\calM}{{\cal M}}
\newcommand{\calN}{{\cal N}}
\newcommand{\calO}{{\cal O}}
\newcommand{\calP}{{\cal P}}
\newcommand{\calT}{{\cal T}}
\newcommand{\calV}{{\cal V}}
\newcommand{\calW}{{\cal W}}

\newcommand{\tilT}{{\tilde{T}}}
\newcommand{\Eint}{E_{\rm int}}

\newcommand {\nav}{\langle n\rangle}
\newcommand {\lav}{\langle \ell\rangle}

\newcommand{\hP}{\hat{P}}
\newcommand{\hPi}{\hat{\Pi}}
\newcommand{\sumn}{\sum_{n=1}^N}

\newcommand{\Ns}{N_{\rm s}}
\newcommand{\Nv}{N_{\rm v}}
\newcommand{\Rs}{R_{\rm s}}
\newcommand{\nnb}{n_{\rm nb}}
\newcommand{\muc}{\mu_{\rm c}}
\newcommand{\munul}{\mu_{0}}
\newcommand{\pc}{p_{\rm c}}
\newcommand{\Co}{C}
\newcommand{\co}{c}

\newcommand{\vecalpha}{\vec{\alpha}}
\newcommand{\vecg}{\vec{g}}
\newcommand{\vecp}{\vec{p}}

\newcommand{\tb}{\tilde{b}}
\newcommand{\tm}{\tilde{m}}
\newcommand{\tA}{\tilde{A}}
\newcommand{\tB}{\tilde{B}}
\newcommand{\tP}{\tilde{P}}
\newcommand{\tbeta}{\tilde{\beta}}
\newcommand{\tgamma}{\tilde{\gamma}}
\newcommand{\tcalM}{\widetilde{\cal M}}
\newcommand{\betast}{{\beta_*}}

\newcommand{\intp}{\int_{-\pi}^{\pi}\frac{\dd p}{2\pi}}
\newcommand{\intpone}{\int_{-\pi}^{\pi}\frac{\dd p_1}{2\pi}}
\newcommand{\intptwo}{\int_{-\pi}^{\pi}\frac{\dd p_2}{2\pi}}
\newcommand{\ointz}{\oint\frac{\dd z}{2\pi{\rm i}}}
\newcommand{\qext}{q_{\rm ext}}

\newcommand{\bO}{{\bf{O}}}
\newcommand{\bR}{{\bf{R}}}
\newcommand{\bS}{{\bf{S}}}
\newcommand{\bT}{{\bf{T}}}
\newcommand{\bn}{{\bf{n}}}
\newcommand{\br}{{\bf{r}}}
\newcommand{\bt}{\mbox{\bf t}}
\newcommand{\half}{\frac{1}{2}}
\newcommand{\thalf}{\tfrac{1}{2}}
\newcommand{\bsA}{\mathbf{A}}
\newcommand{\bsV}{\mathbf{V}}
\newcommand{\bsE}{\mathbf{E}}
\newcommand{\bsT}{\mathbf{T}}
\newcommand{\bse}{\mbox{\bf{1}}}

\newcommand{\invup}{\rule{0ex}{2ex}}
\newcommand{\rmc}{{\rm c}}

\newcommand{\dd}{\mbox{d}}
\newcommand{\p}{\partial}

\newcommand{\la}{\langle}
\newcommand{\ra}{\rangle}

\newcommand{\beq}{\begin{equation}}
\newcommand{\eeq}{\end{equation}}
\newcommand{\bea}{\begin{eqnarray}}
\newcommand{\eea}{\end{eqnarray}}
\def\lsim{\:\raisebox{-0.5ex}{$\stackrel{\textstyle<}{\sim}$}\:}
\def\gsim{\:\raisebox{-0.5ex}{$\stackrel{\textstyle>}{\sim}$}\:}

\numberwithin{equation}{section}   

\thispagestyle{empty}
\title{{\Large {\bf From prelife to life: 
a bio-inspired toy model}\\
\phantom{xxx}}}

\author{{\bf H.J.~Hilhorst}\\[5mm]
{\small Laboratoire de Physique Th\'eorique, UMR 8627}\\[-1mm]
{\small Universit\'e Paris-Sud and  CNRS}\\[-1mm]
{\small B\^atiment 210, 91405 Orsay Cedex, France}\\}

\maketitle

\begin{abstract}
We study a one-dimensional lattice of $N$ sites
each occupied by a mathematical ``polymer,''
that is, is a binary random sequence of arbitrary length $n$,
or equivalently,
a rooted path of $n$ links on an infinite binary tree. 
The average polymer length is controlled by the monomer fugacity $z$.
A pair of polymers on adjacent sites carries a weight factor 
$\omega$ for each link on the tree that they have in common.
The phase diagram in the $z\omega$ plane exhibits a critical line 
$z=\zc(\omega)$.
For $z<\zc(\omega)$ there exists an equilibrium phase with, in particular, a
finite average polymer length. 
We investigate the equilibrium ensemble by transfer matrix and Monte Carlo
methods, paying particular attention to the vicinity of the critical line.
For $z>\zc(\omega)$ the equilibrium is unstable and 
Monte Carlo time evolution brings about
a dynamical symmetry breaking which favors
the evolution of a small selection of polymers to ever greater length.
While of interest for its own sake, this model may also be relevant
to the prelife-to-life transition that has occurred during biological
evolution. 
We compare it to existing models of similar simplicity
due to Wu and Higgs (2009, 2012) and to Chen and Nowak (2012).
\end{abstract}



\newpage

\section{Introduction}
\label{sec_intro}

\cappar Explaining the appearance of life in the prebiotic soup 
billions of years ago is an intriguing but
extremely difficult question. It involves
the statistics as well as the 
physics, chemistry, and biology of interacting macromolecules.
Mathematical models of this prelife-to-life transition, whether simple or
more elaborate, can never do justice to the full complexity of the problem,
but at most shed some light on certain aspects of it.
In this work we present and discuss
a model altogether at the simplistic end of the spectrum.
It may be characterized as a bio-inspired
toy model designed for statistical physicists.

An early statistical model describing the transition from prelife 
to life was formulated by Dyson \cite{Dyson82} decades ago.
At its core there is a bistable Fokker-Planck equation
governing the fraction of monomers that are ``active,''
{\it i.e.} that participate in autocatalytic processes
in the system. The stationary states with the lower and higher fraction
of active monomers are interpreted as prebiotic and alive, respectively. 
Dyson's solution amounts to an application of Kramers' escape rate theory 
\cite{Kramers40}.

Models of similar kind and almost equal simplicity were 
studied by Wu and Higgs \cite{WuHiggs09,WuHiggs12}
and by Chen and Nowak \cite{ChenNowak12}. 
Wu and Higgs \cite{WuHiggs09} consider a set of coupled
rate equations for the concentrations of polymers of given lengths,
without regard for their specific monomer sequence.
Chen and Nowak model a reservoir of
two types of monomers from which an arbitrary number of polymer species may
grow, each species being determined by its length and its monomer sequence.
In all these models ``life'' is identified with the appearance
of an autocatalytic feedback that arises once there is an appreciable
concentration of sufficiently long polymers.

The merits of such models, notwithstanding the
gross simplifications upon which they rely,
have been emphasized by workers in the field of mathematical biology
\cite{Bedau99} and artificial chemistry \cite{BY15}. Dyson 
minimizes the pretensions of his work by stating that
it is ``not intended to be a theory of the origin of life'';  
but stresses that such models may help to ask new questions.\\  

In this work we describe a mixture of species at the level 
of individual polymers which are composed of two types of monomers.
Only two parameters play a role, namely
the monomer fugacity $z$ and a weight factor $\omega$
representing an interaction between polymers.
In the $z\omega$ plane there is a parameter regime corresponding to
an equilibrium state (the prebiotic soup) and another parameter regime
(the state sustaining life) in which the system is unstable and forms
ever longer polymers.
The model shows in particular the possibility of
the emergence of a single or a few dominant polymer species.
Much of this work concentrates on the transition between the two regimes.\\

In section \ref{sec_model} we define the model.
In section \ref{sec_equilibrium} we study its equilibrium state by a
combination of heuristic arguments,
Monte Carlo simulation, and analytic work using the transfer matrix.
In section \ref{sec_timeevolution} we investigate the unstable regime 
by means of heuristic arguments and Monte Carlo simulation.
In our discussion in section \ref{sec_discussion}
we elaborate, in particular, upon the similarities and the differences
between this work and that of references \cite{WuHiggs09}-\cite{ChenNowak12}.
In section \ref{sec_conclusion} we conclude briefly.

\section{Model}
\label{sec_model}

\begin{figure}[t]
\begin{center}
\scalebox{.35}
{\includegraphics{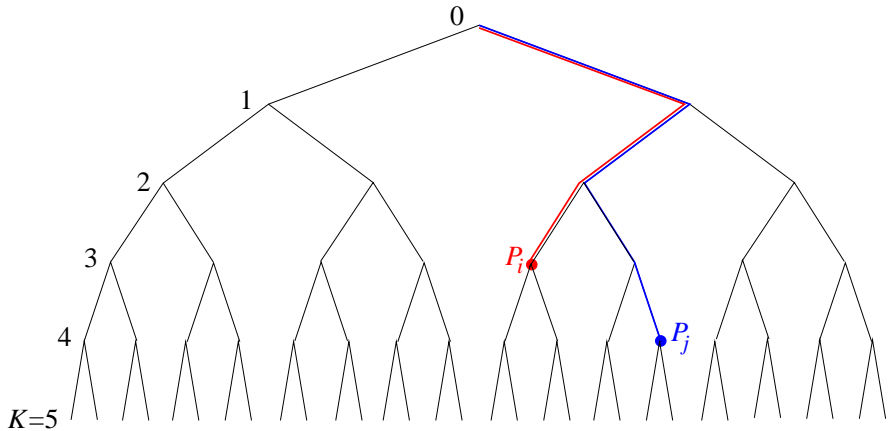}}
\end{center}
\caption{\small A rooted binary tree of depth $K=5$. 
The two rooted paths $P_i$ (red) and $P_j$ (blue)
have lengths 3 and 4, respectively.
They have their first two links in common 
and hence have an overlap $\ell_{ij}=2$.
A rooted path is also called a ``polymer,'' and its links ``monomers.''}
\label{fig_tree}
\end{figure}

Let a one-dimensional lattice have sites $i=1,2,\ldots,N$. 
On each site $i$ lives a mathematical ``polymer'' $P_i$. 
The monomers composing the polymer may be of two different types,
denoted by $+1$ and $-1$. 
We write the polymer variable 
as $P_i=(n_i;p_i^{n_i})$ in which $n_i$ is the number of monomers in $P_i$ 
and where $p_i^{n_i}\equiv(s_{i1},s_{i2},\ldots,s_{i,n_i})$,  
with monomer variables $s_{ik}=\pm 1$, gives the detailed monomer structure
of $P_i$.

Each polymer $P_i$ may be represented by a rooted path 
(that is, a path starting from the origin) on a  binary tree 
(see figure \ref{fig_tree}),
under the convention that a link
going down to the left (down to the
right) corresponds to a monomer of type $-1$ (of type $+1$).
We allow the polymer length $n_i$ to take the values $0,1,2,\ldots,K$,
with $n_i=0$ standing for the absence of a polymer on site $i$, and
where $K$ is a cutoff length that we will send to infinity
at a later stage.\\

By the {\it overlap\,} $\ell_{ij}$  
between two paths $P_i$ and $P_j$
we will mean the number of links on the tree that they have in common.
As is clear from figure \ref{fig_tree},
when the $k$th link is not in common, then the links of indices
higher than $k$, if any, cannot be in common either.%
\footnote{
The overlap 
between $P_i$ and $P_j$ is formally given, therefore, by
$\ell_{ij} = \sum_{k=1}^{\min(n_i,n_j)} \prod_{k'=1}^{k}
\tfrac{1}{2}(1+s_{ik'}s_{jk'})$,
but this expression will not be of help in practice.}
We will also use the more explicit notation
$\ell_{ij} \equiv \ell(p_i^{n_i},p_j^{n_j})$.

Let  $P\equiv (P_1,P_2,\ldots,P_N)$ be a system configuration.
We associate with it an {\it energy\,} $E[P]$ that 
depends on two parameters, namely the monomer chemical potential $\mu$ and
the overlap energy $-\epsilon$.
Explicitly, we set
\beq
E[P] = -\mu \sum_{i=1}^N n_i - \epsilon\sum_{i=1}^{N-1}\ell_{i,i+1}\,.
\label{dEP}
\eeq
The second term on the RHS of (\ref{dEP}) is
one of the simplest ways to introduce a polymer-polymer interaction.
It expresses that in order for one long polymer to catalytically favor
the growth of another one, their two monomer sequences have to be in a precise
relation. 
We note that this second term corresponds to free boundary conditions.

Let $\beta$ stand for the inverse temperature. 
We will employ two alternative variables, the monomer fugacity $z$ and the
``overlap weight'' $\omega$, defined as
\beq
z = \ee^{\beta\mu}\,, \qquad \omega = \ee^{\beta\epsilon}\,,
\label{defzomega}
\eeq
in terms of which the Boltzmann weight of a configuration $P$  becomes
\beq
\ee^{-\beta E[P]} = z^{\sum_{i=1}^N n_i}\,
               \omega^{\sum_{i=1}^{N-1} \ell_{i,i+1} }\,.
\label{dexpEP}
\eeq
We will restrict ourselves to 
$\epsilon>0$, or equivalently to an overlap weight $\omega>1$, 
which favors overlap between neighboring polymers, 
the case $\omega=1$ being the trivial interactionless limit.\\

We will find that in the $z\omega$ plane there is an equilibrium regime
(to be called regime I) where the equilibrium properties of this model are
well-defined in the limit of an infinite cutoff, $K\to\infty$.
This regime is studied in section \ref{sec_reducedTM}. 
 In the remaining regimes (to be called IIa, IIb, and III) there is, for 
 infinite cutoff, no equilibrium state and a dynamic description 
 becomes necessary.
To that end we have endowed this model 
with a standard heat bath Monte Carlo dynamics.
The algorithm allows each polymer $P_i$ to change its length by
addition or suppression of a single monomer at a time,
supposedly due to an exchange with a reservoir of monomers of the two types.

\section{Equilibrium}
\label{sec_equilibrium}

\subsection{Reduced transfer matrix}
\label{sec_reducedTM}

In the equilibrium regime, whose exact location in the $z\omega$ plane has yet
to be determined, the thermodynamic properties of the system
follow from its partition function 
\beq
Z_N(z,\omega) = \sum_P\,\ee^{-\beta E[P]}.
\label{dZN}
\eeq
In terms of $Z_N$ the average polymer length $\la n\ra$
and the average overlap $\la\ell\ra$ between neighboring polymers 
are given by
\beq
\la n\ra   = \frac{1}{N}   \frac{\p\log Z_N}{\p\log z}\,,    \qquad
\la\ell\ra = \frac{1}{N-1} \frac{\p\log Z_N}{\p\log\omega}\,.
\label{dnlav}
\eeq
The calculation of $Z_N$ may be formulated as a transfer matrix problem.
The number $\calK$ of states accessible to the variable $P_i$ 
equals $\calK = \sum_{n_i=0}^K 2^{n_i} = 2^{K+1}-1$,
which would lead to a $\calK\times\calK$ transfer matrix.
We will show now that it is possible at a set of fixed polymer lengths 
$\{n_i\}$ to analytically sum over the polymer configurations
$\{p_i^{n_i}\}$, which then leaves us with a drastically reduced 
transfer matrix of size $(K+1)\times (K+1)$.

Using that $\sum_P = \prod_i\sum_{P_i} = \prod_i\sum_{n_i}\sum_{p_i^{n_i}} $
we have from (\ref{dZN}) and (\ref{dexpEP}),
after suitably arranging the factors,
\bea
Z_N(z,\omega) &=& \sum_{n_1} z^{n_1}\sum_{p_1^{n_1}} 
\sum_{n_2} z^{n_2} \sum_{p_2^{n_2}} \omega^{\ell(p_1^{n_1}\!,p_2^{n_2})}
\sum_{n_3} z^{n_3} \sum_{p_3^{n_3}} \omega^{\ell(p_2^{n_2}\!,p_3^{n_3})} 
\ldots \nonumber\\[2mm]
&&
\ldots
\sum_{n_N} z^{n_N} \sum_{p_N^{n_N}} 
    \omega^{\ell(p_{N-1}^{n_{N-1}}\!,p_N^{n_N})}.
\label{rZN0}
\eea
Because of the free boundary conditions there are no factors of
$\omega$ due to overlap of $p_1^{n_1}$ and $p_N^{n_N}$.
We will now show that the sums on $p_N^{n_N}$, $p_{N-1}^{n_{N-1}}$, \ldots,
$p_2^{n_2}$ in expression (\ref{rZN0})
may be carried out successively in that order.
The sum on $p_N^{n_N}$, which involves $2^{n_N}$ terms, may be rewritten as
\beq
T_{n_{N-1},n_N} \equiv
\sum_{p_N^{n_N}} \omega^{\ell(p_{N-1}^{n_{N-1}}\!,p_N^{n_N})}
= \sum_{\ell=0}^{\min(n_{N-1},n_N)} g_{n_{N-1}n_N}(\ell)\omega^\ell,
\label{xTnma}
\eeq
in which $g_{nm}(\ell)$ is the number of configurations of a
polymer of length $m$ to have an overlap exactly equal to $\ell$
with a given polymer of length $n$.
An elementary calculation with the abbreviation $\nu=\min(m,n)$ leads to
\beq
g_{nm}(\ell) = 2^{\max(m-n,0)}\times\left\{
\begin{array}{ll}
2^{\nu-\ell-1} & 0\leq\ell\leq \nu-1,\\[2mm]
1              & \ell = \nu,
\end{array}
\right.
\label{xgnm}
\eeq
valid for $0 \leq n,m \leq K$.
It is easily verified that $\sum_{\ell=0}^m g_{nm}(\ell)= 2^m$, 
which is the total number of configurations of the polymer of length $m$,
as it had to be. 
Substituting (\ref{xgnm}) in (\ref{xTnma}) and carrying out the sum on
$\ell$ yields
\beq
T_{nm} = 2^{\max(m-n,0)}f(\nu), \quad
f(\nu) = \left\{
\begin{array}{ll}
1, & \nu=0, \\[2mm]
\displaystyle{ \frac{\omega^\nu - 2^\nu}{\omega-2} + \omega^\nu, }
& \nu=1,2,\ldots,K,
\end{array}
\right.
\label{xTnmb}
\eeq
valid for $\omega \neq 2$ (we will leave the special case $\omega=2$ aside). 

Since the result of this summation 
depends on $n_{N-1}$ but not on $p_{N-1}^{n_{N-1}}$, 
we can now repeat the procedure and carry out the sum on $p_{N-1}^{n_{N-1}}$
at fixed path length $n_{N-1}$.
Working our way down through the chain and
still using that the final sum on $p_1^{n_1}$ just yields a factor $2^{n_1}$
we get
\beq
Z_N(z,\omega) = \sum_{n_1}\sum_{n_2}\ldots\sum_{n_N}
2^{n_1} z^{n_1+n_2+\ldots+n_N} T_{n_1,n_2}T_{n_2,n_3}\ldots T_{n_{N-1},n_N}
\,.
\label{rZN1}
\eeq
In terms of the symmetric matrix
\beq
\tilT_{nm} = (2z)^{\frac{n}{2}}\, T_{nm} 
\left( \frac{z}{2} \right)^{\frac{m}{2}} 
= 2^{|n-m|/2} z^{(n+m)/2}f(\nu)
\label{dTS}
\eeq
the partition function (\ref{rZN1}) becomes
\beq
Z_N(z,\omega) = \sum_{n=0}^{K} \sum_{m=0}^{K}  
(2z)^{\frac{n+m}{2}} (\tilT^{N-1})_{nm}\,.
\label{rZN2}
\eeq
This achieves expressing $Z_N$ in terms of a transfer matrix of the
reduced size $K+1$. 

Let $\lambda_0\geq \lambda_1\geq\ldots\geq\lambda_K$ 
be the set of eigenvalues of $\tilT$  and let $\psi^{(k)}$ be the 
normalized eigenvector with
eigenvalue $\lambda_k$, so that we have the decomposition
\beq
\tilT_{nm} = \sum_{k=0}^K \psi^{(k)}_n \lambda_k\, \psi^{(k)}_m.
\label{tilTdecomp}
\eeq
Substitution of (\ref{tilTdecomp}) in (\ref{rZN2}) yields
\beq
Z_N(z,\omega) = \sum_{k=0}^K\, \left| 
\sum_{m=0}^K(2z)^{\tfrac{m}{2}} \psi^{(k)}_m
\right|^2 \lambda_k^{N-1}
\label{rZN3}
\eeq
whence in the limit of large system size
\beq
\log Z_N(z,\omega) = N \log\lambda_0 + {\cal O}(N^0), \quad N\to\infty, 
\label{asptZ}
\eeq
and therefore, with equations (\ref{dnlav}),
\beq
\la n\ra   = \frac{\p\log\lambda_0}{\p\log z} + {\calO}(N^{-1}),    \quad
\la\ell\ra = \frac{\p\log\lambda_0}{\p\log\omega} + {\calO}(N^{-1}), 
\quad N\to\infty.
\label{nllambda0}
\eeq

We now refine the preceding analysis.
Let $Z^{(n,j)}_N(z,\omega)$ be the expression identical to (\ref{rZN1}) 
except for the insertion of an extra factor $\delta_{n,n_j}$ 
in the inner summand;
the ratio $Z^{(n,j)}_N(z,\omega)/Z_N(z,\omega) \equiv \calP^{(j)}_N(n)$
is then the probability that the polymer at site $j$ be exactly of length $n$.
Upon taking $Z^{(n,j)}_N(z,\omega)$ through the same procedure as we did for
$Z_N(z,\omega)$ we find
that in the large $N$ limit and for $j$ sufficiently deep in the bulk, one has
$Z^{(n,j)}_N(z,\omega)/Z_N(z,\omega) \simeq |\psi^{(0)}_n|^2$, 
which is independent of $N$ and of $j$. Hence $\calP^{(j)}_N(n)$
has the limit distribution
\beq
\calP(n) = |\psi^{(0)}_n|^2, \qquad n=0,1,\ldots,K, \qquad N\to\infty.
\label{xcalPn}
\eeq
The normalization of $\psi^{(0)}$ implies that of $\calP$ and
{\it vice versa}.

Equations (\ref{asptZ}) and (\ref{xcalPn}), therefore, relate
the quantities of physical interest $\nav, \lav$, and $\calP(n)$, to
the largest eigenvalue $\lambda_0$ and its eigenvector $\psi^{(0)}$.
To find $\lambda_0$ and $\psi^{(0)}$ we will have recourse to numerical 
techniques in section \ref{sec_numericalTM}. 
Before applying these, however, 
we present in section \ref{sec_heuristics1}
a heuristic argument that will establish the boundary delimiting
the equilibrium regime (regime I) in the $z\omega$ plane,
and in section \ref{sec_equilibriumMC}
some Monte Carlo results that illustrate the equilibrium behavior 
of the polymers.

\subsection{Heuristics: Phase diagram}
\label{sec_heuristics1}

\begin{figure}[t]
\begin{center}
\scalebox{.45}
{\includegraphics{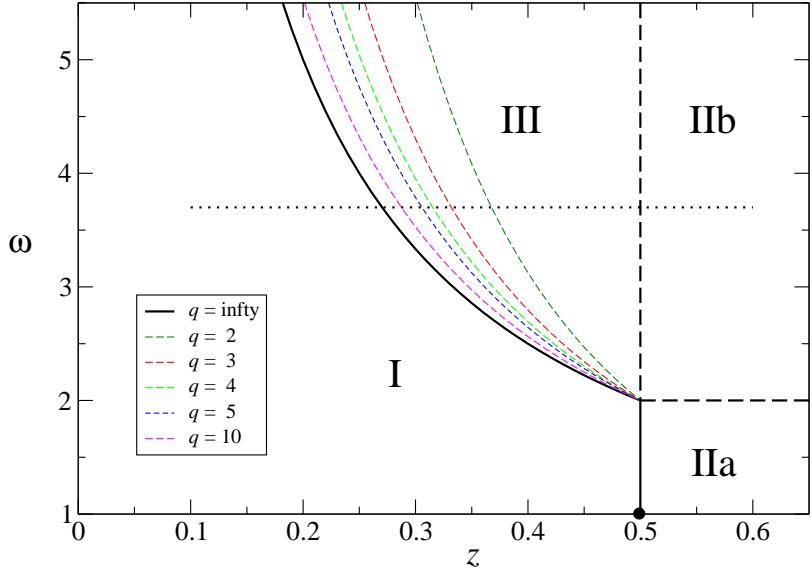}}
\end{center}
\caption{\small
Phase diagram in the $z\omega$ plane.
In the limit of cutoff $K\to\infty$ there appear three regimes,
separated by solid or dashed black lines.
In regime I a true equilibrium state exists with a finite average polymer
length $\nav$;
in regimes II and III the equilibrium is unstable 
and the polymer length may grow without limit.
In regime II ($z>\tfrac{1}{2}$) this unbounded growth is due to
the sole pressure of the monomer fugacity $z$;
the distinction between IIa and IIb is briefly discussed 
at the end of section \ref{sec_simulation}.
In regime III the unbounded growth is due
to the {\it combined\,} effect of the monomer fugacity $z$ {\it and\,} 
the polymer-polymer interaction $\omega$; this is the analog of autocatalytic
self-replication in other models.
The curved part of the boundary between regimes I and III is given by
$z\omega=1$. The colored dashed curves are given by $\omega = (2z^q)^{-1/(q-1)}$,
for $q=2,3,4,\ldots$ and are interpreted in section \ref{sec_heuristics2}.
The dot on the $z$ axis indicates the phase transition point in the
trivial interaction\-less case, $\omega=1$.
The simulations in this work are at fixed $\omega=3.7$ and for varying $z$,
that is, along the thin dotted horizontal line.}
\label{fig_zomega}
\end{figure}

We expect that for small enough $z$ the average polymer length 
and other physical quantities will 
have finite values that tend to 
well-defined limits when the cutoff $K$ is sent to infinity.
But we also expect that for large enough $z$ the typical polymer
will be as long as is allowed by the cutoff $K$.
Hence for $K\to\infty$ there must be in the $z\omega$
plane a phase boundary $z=\zc(\omega)$ between these two regimes. 
We present now a heuristic argument that determines this phase boundary
by balancing entropy against energy.

In the trivial interactionless case, $\omega=1$,
when the length of a polymer is increased from $n$ to $n'$,
its Boltzmann weight acquires a factor $(2z)^{n'-n}$,
the entropic coefficient $2$ being due to the two 
types of monomers that are possible at each unit length increase.
This establishes that for $z<\tfrac{1}{2}$ the distribution of the
polymer length $n$ will decay exponentially,
whereas for $z>\tfrac{1}{2}$ the polymers will grow 
until stopped by the cutoff $K$.
The heavy dot on the $z$ axis in figure \ref{fig_zomega} 
separates the two regimes.

We now consider how this picture is changed in the case of
an overlap weight $\omega>1$. 
Let us first suppose that all polymers have a length of order $n$ 
where $n$ is large. 
If a polymer of length $n$ is forced to be identical to a 
neighboring polymer
of at least the same length, its entropy change is $-n\log 2$ (it goes down)
and its energy change is $-\epsilon n$ (it also goes down).
Hence its Boltzmann weight acquires a factor $(\omega/2)^n$.
Making neighboring polymers {\it of given lengths\,}
identical is favorable when this factor exceeds unity,
that is, when $\omega>2$.
Next we determine under which conditions the polymers will 
satisfy the prerequisite of having large lengths.
Suppose a site is occupied by a long polymer of length $n$.
Placing on the site next to it an identical polymer will multiply the
Boltzmann weight by a factor $z^n\omega^n$. Hence for
$z\omega>1$ it will be favorable for $n$ to become large.

These arguments together define the equilibrium regime (regime I) of the phase 
diagram: it is located to the left of the boundary 
$\zc(\omega)=\min(\tfrac{1}{2},\omega^{-1})$,
shown as a solid black line in figure \ref{fig_zomega} and consisting
of a straight vertical segment and a curved part.
When this bounday is approached from the left, the equilibrium average $\nav$
diverges. There is, however, a difference between entering regime IIa and
entering regime III.
When the boundary with regime IIa is approached, $\lav$ remains finite:
the polymers can grow to infinity under the sole influence of the monomer
fugacity $z$ and their entropy gain associated with having greater length.
By contrast, when the boundary with regime III is approached, 
the argument of the preceding paragraph indicates that $\lav$ diverges along
with $\nav$. 
We identify regime III as the regime of greatest interest:
it is characterized by interaction mediated unlimited polymer growth.
In this regime the interaction between identical neighboring polymers
produces consequences similar 
to the autocatalytic effects incorporated in other models 
\cite{Dyson82,WuHiggs09,WuHiggs12,ChenNowak12}.
We will analyze this phenomenon in detail in section \ref{sec_timeevolution}
by investigating the behavior of this model along the dotted line in figure 
\ref{fig_zomega}, that is, as a function of $z$ 
at a fixed value of $\omega$.

In spite of the heuristic nature of these arguments,
we believe on the basis of what will follow below that the
results so obtained are exact. 

\subsection{Monte Carlo simulation in equilibrium}
\label{sec_equilibriumMC}

We have performed standard heat bath Monte Carlo dynamics,
allowing each polymer $P_i$ to change its length by
addition or suppression of a single monomer at a time,
attributable to exchange with a reservoir of the two types of monomers.
In regime I this dynamics is guaranteed to
reproduce the equilibrium statistics of the model. 

\begin{figure}[t]
\begin{center}
\scalebox{.45}
{\includegraphics{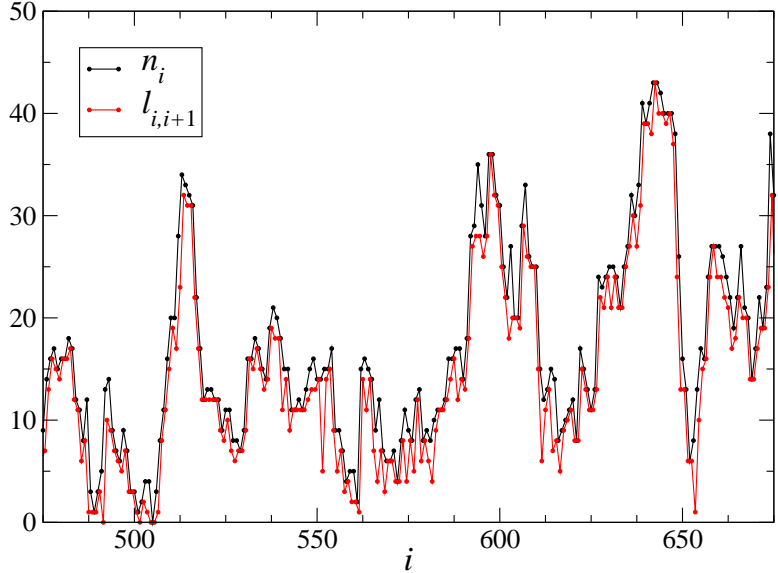}}
\end{center}
\caption{\small A typical local
equilibrium configuration of the $n_i$ and $\ell_{i,i+1}$
at overlap weight $\omega=3.7$ and 
for a monomer fugacity $z=0.26900$, below the critical point $\zc=0.27027$. 
The $\ell_{i,i+1}$ have been plotted at the half-integer
coordinates $i+\tfrac{1}{2}$.
There is very strong correlation between the $\ell_{i,i+1}$ and their 
neighboring $n_i$. For an initial configuration without polymers (all
$n_i=0$), the equilibration time is of the order of $t=5\times10^5$
sweeps through the lattice.
}
\label{fig_stable26900}
\end{figure}

We carried out a simulation at an arbitrarily fixed value $\omega=3.7$ 
while choosing $z=0.26900$
closely below the (at this stage still presumed) critical point 
$\zc = 1/\omega = 0.270270...$. 
Figure \ref{fig_stable26900} shows a typical equilibrium configuration of 
the polymer lengths $n_i$ as a function of the site
index $i$ for a portion of a larger system.
It also shows the nearest-neighbor overlaps $\ell_{i,i+1}$.
The strong correlation between these two ``profiles'' 
shows that in order for large fluctuations to arise there has to be
sufficient overlap between neighboring polymer pairs, 
in agreement with the heuristic argument of the preceding subsection. 

\begin{figure}[t]
\begin{center}
\scalebox{.45}
{\includegraphics{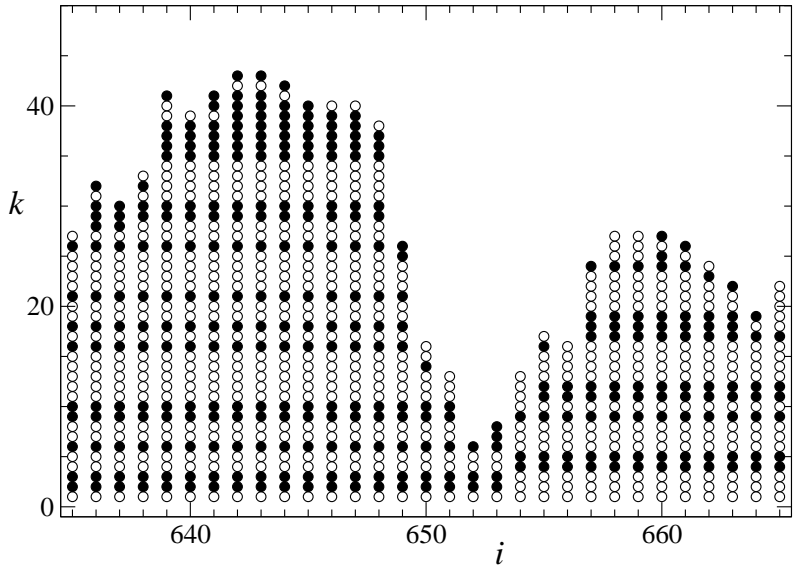}}
\end{center}
\caption{\small Lattice representation of the full monomer sequences 
of the polymers on sites 635 through 665 for the same configuration that led
to figure \ref{fig_stable26900}. 
Site $(i,k)$ in this figure is marked by an
open or a closed circle for monomer variable $s_{ik}=1$ or $s_{ik}=-1$,
respectively.
The overlaps $\ell_{i,i+1}$ may be read off from this representation.
In the figure they attain a minimum value equal to 1 between
the sites $i=653$ and $i+1=654$.
}
\label{fig_polymers}
\end{figure}

The profiles are only projections of the full phase
space configuration in that they
hide the underlying structure of the polymers as sequences of two types of 
monomers. As an illustration of the monomer structure
we represent in figure \ref{fig_polymers}
the specific monomer sequences of the polymers on  
sites $i=635$ through $i=666$ corresponding to figure \ref{fig_stable26900}. 
Figure \ref{fig_polymers} shows that there are dips in the overlap that
coincide with dips in the polymer lengths.

We will be especially interested in the approach of the critical point,
$z=\zc$. Figure \ref{fig_divnl} shows as open black circles the
Monte Carlo results for $\nav$ as a function of $\zc-z$ 
on a lattice of $N=960$ sites. These data result from averaging
over a succession of $10^{10}$ attempted moves, that is,
over $1.04\times 10^7$ sweeps through the lattice.
They have error bars less than their symbol size.
We consider them as preliminary to the transfer matrix results 
for $\nav$, to be discussed in the next subsection.

\subsection{Transfer matrix based numerical analysis}
\label{sec_numericalTM}

We have not been able to diagonalize the reduced transfer matrix 
$\tilT$ of equation (\ref{dTS}) analytically.
However, exploiting it numerically is easy. 
At the same fixed value $\omega=3.7$ of the interaction we
have studied the average polymer length $\nav$ and average overlap $\lav$ 
for varying monomer fugacity $z$, 
that is, along the thin dotted line in figure \ref{fig_zomega}.
We proceeded by numerically finding the largest eigenvalue $\lambda_0$ and 
corresponding eigenvector $\psi^{(0)}$ of 
$\tilT$ and we obtained $\nav$ and $\lav$ from it by numerical
differentiation. At each value of $z$
the procedure was carried out for increasing values of the
cutoff until convergence was obtained.

\begin{figure}[t]
\begin{center}
\scalebox{.45}
{\includegraphics{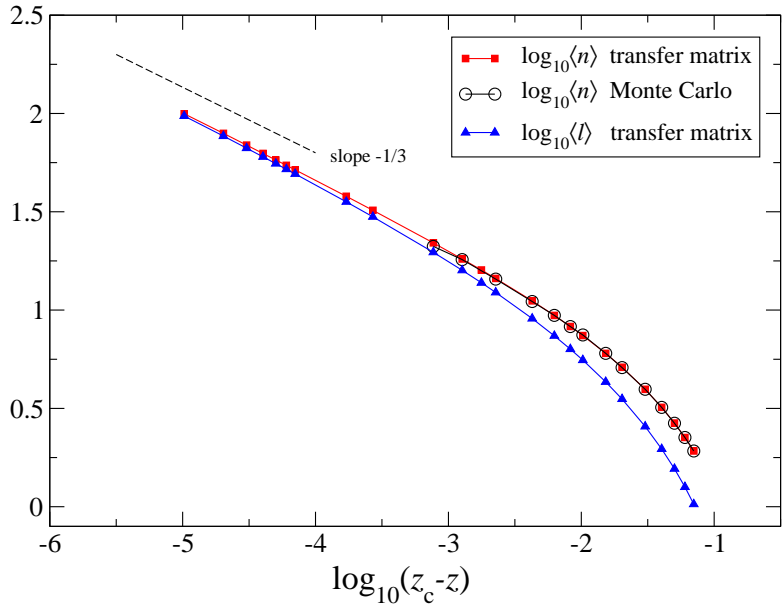}}
\end{center}
\caption{\small
Divergence of $\nav$ and $\lav$ as the monomer fugacity $z$ approaches 
the critical point $\zc$ from below at overlap weight $\omega=3.7$.
This log-log plot, together with the arguments presented in the text, 
strongly suggests the asymptotic behavior $\nav \sim \lav \sim (\zc-z)^{-1/3}$.
The Monte Carlo data were obtained as described in section 
\ref{sec_equilibriumMC}, the transfer matrix results as in section 
\ref{sec_numericalTM}. The two methods fully agree,
the latter being clearly more powerful.}
\label{fig_divnl}
\end{figure}

A critical point appears at a location fully compatible with the heuristic
prediction $z=\zc=\omega^{-1}=0.270270...$.
We have therefore plotted our numerically exact results in figure 
\ref{fig_divnl} as a function of $\zc-z$.
Our data point closest to $\zc$ is at $z=0.27026$;
it has $\nav = 99.6$ and required a cutoff $K \gsim 300$.
This figure shows that as $\zc$ is approached, the data points follow
asymptotically a straight line, thereby lending support to
our assumed value of $\zc$.
The figure strongly suggests, furthermore, 
that $\lav$ has the same asymptote as $\nav$.
This means that upon approach of criticality
neighboring polymers are correlated over almost their full length:
they can only grow coherently.

The log-log plot of figure \ref{fig_divnl} implies that
\beq
\la n\ra \sim (z_{\rm c}-z)^{-\alpha},  \qquad \zc-z\to 0.  
\label{loglogn}
\eeq
If one considers likely that the exponent is a simple fraction, then
the figure is a strong indication that $\alpha=1/3$.
An explanation of this exponent value appears if one views
the succession of polymer lengths $n_1,n_2,\ldots,n_N$
as the height variables of a one-dimensional interface near a wall 
(the latter represented by all $n_i=0$)
in a potential that increases with the distance from the wall.
In such models the interface width is known to diverge 
with the $1/3$ power of the inverse potential strength 
\cite{vanLeeuwenHilhorst81}.
We elaborate on this approximate analogy in Appendix \ref{sec_appendix}.\\

\begin{figure}[t]
\begin{center}
\scalebox{.45}
{\includegraphics{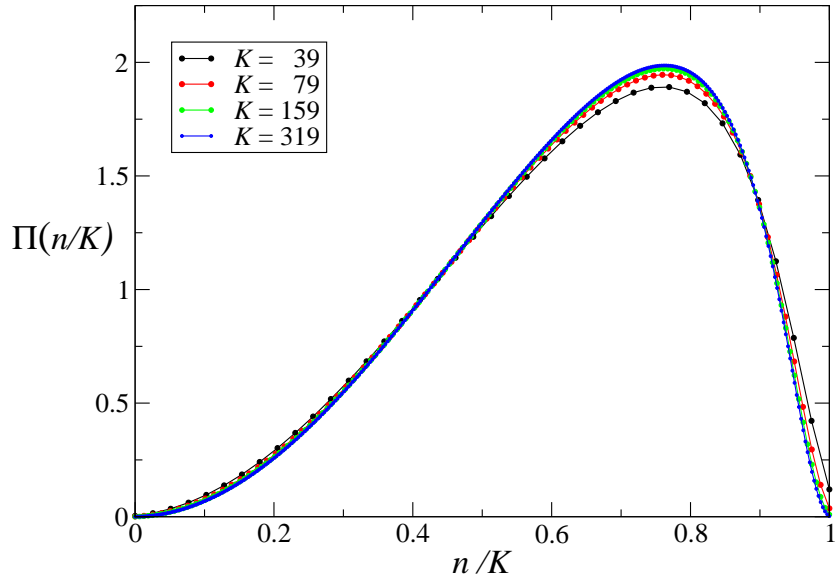}}
\end{center}
\caption{\small
Critical probability distribution $\Pi(n/K)$ of the scaled polymer length 
$n/K$ at $\omega=3.7$ and for different values of the cutoff length $K$.
As $K$ increases, the scaled distribution appears to tend to a limit.}
\label{fig_scalingPk}
\end{figure}

We have, next, studied the polymer length distribution $\calP(n)$.
As expected, this distribution is concentrated near the origin $n=0$
for $z<\zc$ and near the cutoff $n=K$ for $z>\zc$.
Its behavior exactly at the critical point $z=\zc$ is interesting.
We have determined it from the largest eigenvector 
$\psi^{(0)}$ of the transfer matrix with the aid of equation (\ref{xcalPn})
for different values of the cutoff $K$.
A data collapse is obtained by the scaling
\beq
\calP(n) = \frac{1}{K}\, \Pi\left(\frac{n}{K}\right) 
\label{calPnscaling}
\eeq
and shown in figure \ref{fig_scalingPk}.
We observed that even for very small nonzero $|z-z_{\rm c}|$ the curve $\Pi(x)$ 
develops discontinuities of slope near both ends of its interval.
We consider this as
another confirmation of the exactness of the location of the critical point 
at $\zc=1/\omega$. 

As a consequence of equation (\ref{calPnscaling}) the average polymer length
at the critical point is proportional to the cutoff,
\beq
\la n\ra_{\rm c} \simeq aK,  \qquad K\to\infty,
\label{navc}
\eeq
with an estimated coefficient $a\equiv \int_0^1 \dd x\,x\,\Pi(x)\approx 0.639$.


\section{Time evolution}
\label{sec_timeevolution}

\subsection{Cutoff and time evolution}

Once we set $K=\infty$ there exists in regimes II and III neither an
equilibrium state nor a nonequilibrium steady state.
When starting with a collection of polymers of finite length
(or, for simplicity, in the state of zero polymers, $n_i=0$
for all $i=1,2,...,N$),
we may ask what happens under the Monte Carlo dynamics
in these unstable regimes.
We certainly expect formation of polymers of ever increasing length,
and it is of interest to investigate the asymptotic behavior of this
process. 
There is no easy way to analytically answer these questions
and we will therefore have recourse to simulation and heuristic arguments. 

\begin{figure}[t]
\begin{center}
\scalebox{.45}
{\includegraphics{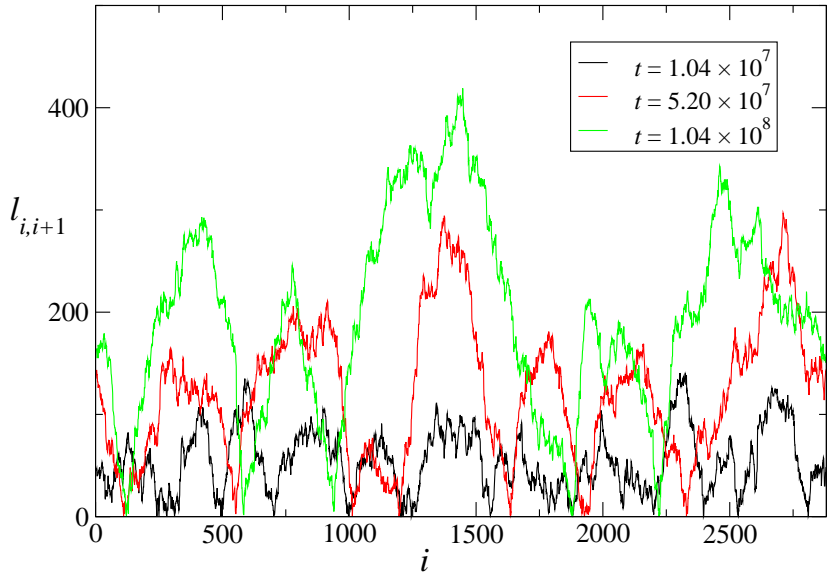}}
\end{center}
\caption{\small The overlaps $\ell_{i,i+1}$ as a function of the lattice site 
coordinate $i$ at overlap weight $\omega=3.7$ and monomer fugacity
$z=0.27040$, that is, just above the critical point $z_{\rm c}=0.27027$.
The profile is for a system of $N=2880$ sites at three different times.
The correlation length of the profile is seen to grow with time.
}
\label{fig_instab27040}
\end{figure}

\begin{figure}[t]
\begin{center}
\scalebox{.45}
{\includegraphics{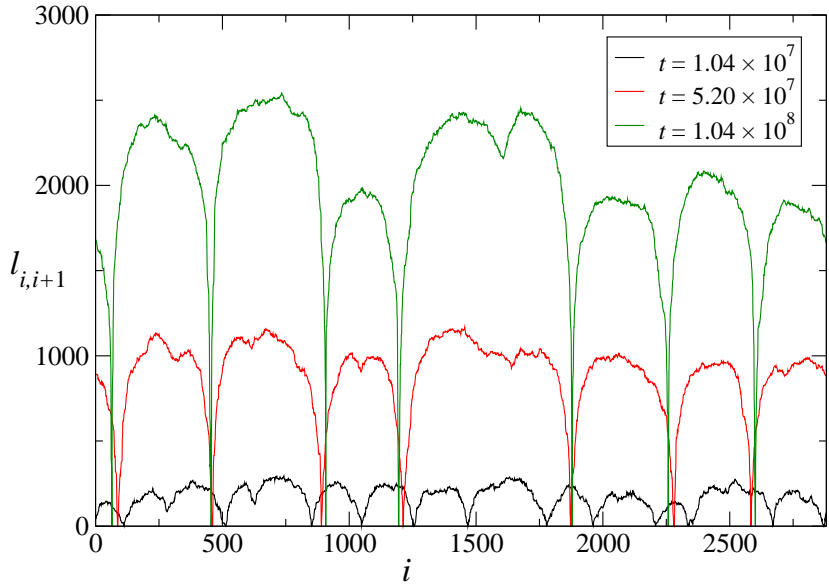}}
\end{center}
\caption{\small The overlaps $\ell_{i,i+1}$ at overlap weight
$\omega=3.7$ and monomer fugacity
$z=0.27100$ in a system of $2880$ sites at three different times.
Comparison of the profiles at the later two times shows that the growth of the
typical block size has come to a standstill.
}
\label{fig_instab27100}
\end{figure}

\begin{figure}[t]
\begin{center}
\scalebox{.45}
{\includegraphics{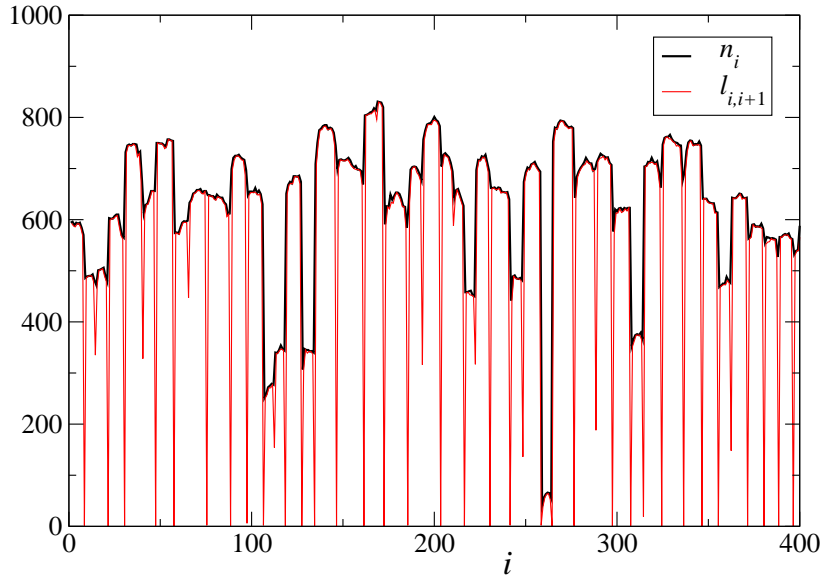}}
\end{center}
\caption{\small The polymer lengths $n_i$ (black) and overlaps 
$\ell_{i,i+1}$ (red) at overlap weight $\omega=3.7$ and for monomer
fugacity  $z=0.3$ in a system of $N=400$ sites, 
as observed at time $t=2.5\times 10^5$.}
\label{fig_instab30000}
\end{figure}

\subsection{Heuristics: Growth in the unstable regime}
\label{sec_heuristics2}

We first present a heuristic argument
that tries to describe the time evolution in regime III.
By a ``block'' we will mean an interval of sites occupied by sufficiently
long strongly overlapping polymers 
(strong overlap meaning that $\ell_{i,i+1}$ is typically close to
its maximum possible value $\min(n_i,n_{i+1})$),
whereas the overlap with the two polymers on the sites just
outside the interval is negligible.

We consider an idealized block 
of $q$ sites $i+1,i+2,\ldots,i+q$ occupied by $q$ identical
polymers of length $n$,
so that all nearest neighbor overlaps are also equal to $n$.
Let moreover the polymers at the end sites of this interval
have zero overlap with their neighbors just outside the interval.
Suppose now that we increase each of the $q$ polymers by a single link, 
the same one on all $q$ sites. This will multiply the 
weight of this set of polymers
by the factor $2z^{q}\omega^{q-1}$, the coefficient $2$ being due to the two
possible choices for the new link.
This factor is larger than unity when the block size%
\footnote{Our terminolgy will be to speak of the {\it size\,} of a block,
as opposed to the {\it lengths\,} of the polymers that constitute it.}
$q$ is larger than a ``correlation length'' $\xic$ given by
\beq
\xic(z,\omega) \equiv \frac{\log(\omega/2)}{\log(z\omega)}\,.
\label{ineqX}
\eeq
For $q>\xic$ the polymers on this interval can 
grow collectively without limit; for $q<\xic$ no such growth is possible. 
Hence $\xic$ is a dynamically required 
minimum correlation length.

For $z_{\rm c}<z<\tfrac{1}{2}$ the correlation length $\xic$ goes down
smoothly from infinity to unity.
Based on equation (\ref{ineqX}) we may in the phase diagram construct the
curves $\xic=q$, which leads to the set of curves
\beq
\omega=(2z^q)^{-1/(q-1)}, \qquad q=1,2,\ldots,
\label{omegazq}
\eeq
represented by colored dashed lines in figure \ref{fig_zomega}. 
According to the argument above,
in the subregion of regime III between the curves labeled $q-1$ and $q$
the average polymer length in a block
can grow only for a block size at least equal to $q$.
The vertical black dashed line is the curve for $q=1$.
To its right
the polymers no longer need their interaction to grow and a growing polymer
will eventually follow its own unique path in the binary tree.

In the real system juxtaposed blocks will interact, which
augments the approximate nature of the preceding argument. 
We expect, nevertheless, 
to see the effects discussed above at least qualitatively
in the Monte Carlo simulations.

\subsection{Simulation results}
\label{sec_simulation}

We have simulated the polymer growth in regime III,
continuing along the same line $\omega=3.7$ (see figure \ref{fig_zomega}) that
was studied in section \ref{sec_numericalTM},
and starting from an initial state without any polymers. 
Our results involve lattices of different sizes $N$ and in each case 
the time variable $t$ will stand for the physical time, that is,
the average number of update attempts per lattice site.
We imposed periodic boundary conditions, mainly in order to obtain better
estimates of bulk quantities,
and chose the cutoff large enough
(often $K=10\,000$) so that it was never attained
during the time interval of the simulation.

Our principal finding is that above the critical point, $z>\zc$, 
there occurs a dynamical symmetry breaking:
the system forms blocks of finite size that, at least initially, 
become larger due to a coarsening process.
As soon as a block size exceeds the minimum required value
$\xic(z,\omega)$, the average polymer length in the block 
(the ``block profile'') starts growing
quasi-independently of its neighboring blocks.
Under the biological interpretation such a block 
represents a set of self-replicating
molecules of a selected species.
We will now show this in detail.

In figure \ref{fig_instab27040} the monomer fugacity $z=0.27040$ is 
just above the critical point $\zc=0.27027$.
This figure shows the profile of the overlaps $\ell_{i,i+1}$ as a function of
the lattice site coordinate $i$ in a system of $N=2880$ sites and
for three different times.%
\footnote{Just as is figure \ref{fig_stable26900},  
the profiles of the $n_i$ very nearly coincide with 
those of the $\ell_{i,i+1}$. We therefore do not show them in figure
\ref{fig_instab27040}.}
One clearly distinguishes blocks in the above defined sense:
large size intervals of considerable overlap are
separated by narrow, almost point-like, intervals of close-to-zero overlap. 

As time goes on two things happen, namely (i) the
amplitude of the profile fluctuations increases; 
and (ii) there is a coarsening causing the typical block size, 
that we will denote by $\xi(t)$, to increase with time. 
We will also refer to $\xi(t)$ as the system's time dependent
{\it correlation length\,}. 
This length $\xi(t)$ should be compared with the value $\xic=1282$, obtained
from equation (\ref{ineqX}). 
In figure \ref{fig_instab27040} 
the simulation time $t=1.04\times 10^8$ 
is not long enough for $\xi(t)$ to reach this minimum value.
As a consequence the three curves are still in what should be called the
critical regime: each of them resembles the equilibrium profile of figure
\ref{fig_stable26900}. At these values of $z$ and $\omega$
longer simulation for a larger lattice would be needed
to take the system out of the critical region. 

In figure \ref{fig_instab27100} the overlap weight still has the same value
$\omega=3.7$ but the monomer fugacity $z=0.27100$
is further above the critical point $\zc=0.27027$ and the growth of
the profile, shown at the same three times as in
figure \ref{fig_instab27040}, is much faster. 
The black curve, taken at the earliest time 
$t=1.04\times 10^7$, has twelve clearly marked minima at or close to zero,
which corresponds to a correlation length that we may roughly estimate as
$\xi(t)=2880/12=240$.
At the later time $t = 5.20 \times 10^7$ the red curve 
shows that the system has coarsened to only seven such minima,
whence a correlation length that has gone up to $\xi(t) = 2880/7=411$.
However, at still later times, the green curve shows that 
the coarsening appears to have stopped, with a  
correlation length frozen%
\footnote{We cannot exclude mathematically that  
for much later times the coarsening still continues
at some exponentially small rate; however, for all practical
purposes it has come to an absolute stop.}
at the asymptotic value $\xi(\infty) = 411$.
For comparison, equation (\ref{ineqX}) gives $\xic=228$ for this
pair $(z,\omega)$.
We observe here that in the actual system the block sizes are distributed in
an interval that extends roughly from $\xic$ to $2\xic$. 
This is easily understood:
blocks smaller than $\xic$ cannot grow and blocks larger than $2\xic$ 
gain entropy by splitting up into two blocks both larger than $\xic$.

On the asymptotic time scale the profiles in the different block
profiles appear to grow linearly with time. 
Growth speeds are somewhat block size dependent, being larger for the
larger blocks.
The largest size block in figure \ref{fig_instab27100}, 
namely the one extending between approximately $i=1200$ and $i=1900$,
shows a tendency to split in two, which causes it growth to
somewhat slow down. 

Figure \ref{fig_instab30000} is still for $\omega=3.7$ but
was obtained for monomer fugacity $z=0.3$, 
well above the critical point $\zc=0.27027$.
For this pair $(z,\omega)$ equation (\ref{ineqX}) yields $\xic=5.9$.
The polymer growth is much faster than near the critical point
and the data were taken at time $t=2.5\times 10^5$.

The red curve represents the $\ell_{i,i+1}$ profile; in this figure
we have again shown the
$n_i$ profile, which differs from the $\ell_{i,i+1}$ near the block
boundaries. 
Several of the blocks are subject to splitting attempts,
which makes the determination of the typical block size somewhat ambiguous.
By taking into account that the $\ell_{i,i+1}$ profile has
about $33$ zeros or near-zeros we arrive at the estimate
$\xi(\infty) = 400/33=12.1$.
The actual block sizes are again distributed 
in a range going from $\xic$ to about $3\xic$,
again confirming the role played by $\xic$: growing blocks respect this
minimum size condition.
The typical block profile grows again linearly with time.
Clearly some of the smaller blocks, of sizes around or below the minimum size,
do not grow well.\\

For completeness we briefly discuss regimes IIa and IIb.
When $z$ is further increased, 
the dynamically required minimum correlation length $\xic$
crosses unity at $z=\tfrac{1}{2}$ and the system enters regime IIb. 
For $z>\tfrac{1}{2}$ each polymer can grow independently of its neighbors.
This is ``disordered growth'' in the sense that, contrary to what 
happens in regime III, no specific polymer types are selected and dominate.
In regime IIb blocks of two (or more rarely a few) 
polymers are observed to grow coherently
until at some point in time they break up, after which
each polymer grows independently of its neighbors. 
This entails that $\lav$ saturates at a finite value:
the $\ell_{i,i+1}$ decouple from the $n_i$\,.
The typical break-up time increases with $\omega$ and decreases with $z$.

The dynamical scenario in regime IIa is qualitatively the same as in IIb.
The two regimes distinguish themselves  
by the behavior of the limit $K\to\infty$:
in regime IIb this limit is accompanied by 
both $\nav\to\infty$ and $\lav\to\infty$, whereas in regime IIa
we have $\nav\to\infty$ but $\lav$ remains finite. The observed fact that 
dynamically $\lav$ remains finite in both regimes
shows that the limits $K\to\infty$ and $t\to\infty$ do not commute. 

\section{Discussion}
\label{sec_discussion}

Biological evolution necessarily involves a
transition from prelife to life. 
With\-in the context of simplified models 
``prelife'' is considered as characterized by 
the spontaneous growth of long polymers 
due to the addition of single monomers one at a time, 
whereas ``life'' is characterized by the self-replication, 
or autocatalytic  polymerization, of 
certain polymer species.
Models of interest are those in which 
a tuning of parameters may take us from a prelife phase to a life phase. 
After having in the preceding sections exposed and analyzed our model 
we will discuss here some of its similarities and differences with, 
specifically, the work of Wu and Higgs 
\cite{WuHiggs09,WuHiggs12}
and of Chen and Nowak \cite{ChenNowak12}.\\

Wu and Higgs \cite{WuHiggs09} 
study a system of polymers of which they ignore the 
nucleotide sequences, keeping track only of their lengths.
They formulate a set of coupled rate equations 
for the concentrations of polymers of given length.
A nonlinear term in the equations represents the autocatalytic effect
responsible for self-replication; 
it becomes operative only once spontaneous 
polymerization in the prelife state has led to
the appearance of polymers exceeding a fixed threshold length. 
The rate equations then have two stable stationary states
of which the one with the lower (higher) concentration of long polymers
is interpreted as the ``prelife'' (``life'') state.

In a subsequent two-dimensional non-well-mixed version of their model
Wu and Higgs \cite{WuHiggs12} study spatial fluctuations. 
They find that the prelife state is metastable and
that the appearance of life is a one-time local stochastic event. 
It is, essentially, analogous to the process, well-known in statistical
physics, of crossing a nucleation barrier in the Ising model.\\

Ignoring the specificity of the 
monomer sequences precludes observing the phenomenon of selection. 
The work by Chen and Nowak \cite{ChenNowak12}
does describe such a selection.
These authors consider a reservoir of
two types of monomers from which an arbitrary number of polymer species 
$p=(n,p^n)$ may grow,
\footnote{The notation here is as in section \ref{sec_model}: 
$p^n$ is a sequence of $n$ binary variables.} 
that is, their phase space has the same binary tree structure as ours.
In order to represent the complexity of the actual chemical kinetics,
the authors 
introduce randomly fixed spontaneous growth rates, 
controlled by a parameter $s$,
and randomly fixed replication rates, 
controlled by a parameter $r$,
the latter rates being intended to mimick the
autocatalytic effects. 
Chen and Nowak then set up rate equations for the time
evolution of the species concentrations $x_p(t)$.
All species grow and replicate independently of one another,
apart from a collective scaling imposed by a depletion term
that keeps the total concentration fixed. 

Chen and Nowak's model has polymerization but no depolymerization reactions. 
There is, therefore, no such thing as detailed balancing,
nor a thermodynamical equilibrium.
The system evolves towards a stationary state%
\footnote{
The authors use the term ``equilibrium state'' for what 
in statistical physics is usually called a (nonequilibrium) 
stationary state.} 
which, depending on the value of the replication parameter $r$,
may either contain a mix of species of different lengths and compositions, 
or be strongly dominated by the abundance of a single species, 
selected by the random landscape. Between the two regimes
there is no sharply defined phase transition point
in the sense of statistical mechanics; 
however, on the $r$ axis 
a narrow ``critical interval'' separates prelife from life.\\ 

By comparison,
our model provides a description in terms not of concentrations but 
of individual polymers.
It is governed not by rate equations but by a multidimensional
master equation, implemented by Monte Carlo dynamics.
Our two parameters, the monomer fugacity $z$ and the weight
factor $\omega$ that represents the interaction strength, 
play very approximately the same role as the parameters 
$r$ and $s$, respectively, in reference \cite{ChenNowak12}.

The most distinguishing feature of the present model is 
the presence of a polymer-polymer interaction,
albeit one of a very elementary kind, 
which expresses that in order for one long polymer to
catalytically favor the creation of another one,
their two monomer sequences have to be in a precise relation.
A consequence of this interaction is that polymer growth is possible only as a  
collective phenomenon. 
It leads to the growth 
of a limited set of preferred species, the selection 
being determined by a random dynamical symmetry breaking.
This selection, which occurs in a regime of the $z\omega$ plane that we
identified as regime III,
illustrates a theme stated in reference \cite{ChenNowak12},
namely that prelife allows coexistence of many species, but that life leads to
competitive exclusion.

The main conclusion of this work is that 
selection of certain species at the cost of others
may be the result of interaction between species; 
and that a random parameter landscape, even though chemically
realistic, is not a necessary ingredient.
We believe we have provided a complementary way to look at the questions
studied 
in references \cite{Dyson82,WuHiggs09,WuHiggs12,ChenNowak12}
and hope that this model,
or future variants of it, will be of interest to statistical physicists.

A further point deserves mention.
The polymers in this work grow without limit.
It would be easy to introduce a smooth cutoff by including 
a depletion process.
We consider, however, this infinite growth as the germ of open-endedness
that in later work begs to be implemented by cooperative events
of greater complexity between the polymers.

\section{Conclusion}
\label{sec_conclusion}

The transition from prelife to life, that is, from the prebiotic soup  
to an environment dominated by specific self-replicating long
polymers, is an extremely complicated question.
It has nevertheless led to a few very simple models in mathematical biology.
These in turn have inspired us to construct a toy model meant to be of interest 
to statistical physicists. 

In this model, each site of a one-dimensional lattice is occupied by a
polymer, that is,
a binary sequence of monomers of variable length.
The system evolves in time by addition or suppression of single monomers.
Its one-dimensional structure, certainly artificial,
has served us to check the Monte Carlo data against analytic results. 

The model has an equilibrium phase, in the sense of statistical mechanics,
and a ``phase,'' in a more general sense, that is neither stationary
nor evolves towards a stationary one, 
and is dominated by specific selected polymers.
We view this latter phase as the rudimentary precursor of an open-endedly
evolving system. The existence of this ``life phase''
is not due to an explicitly incorporated autocatalytic term 
as in earlier work \cite{WuHiggs09,WuHiggs12,ChenNowak12},
but is the consequence of an interaction at the monomer level 
between different polymer species. 
The prelife-to-life transition is a transition between these two phases.

We have not explored all aspects of this model and
many further questions could be asked.
Other models of the same kind also seem worthy
of being developed and studied.
Preliminary results by ourselves indicate that a mean-field 
(``well-mixed'') version 
of this model has essentially the same properties as those
found here in one dimension.
We leave the study of these and other extensions to future work.


\appendix

\section{Appendix: The $\{n_i\}$ viewed as an interface}
\label{sec_appendix}

We consider in this Appendix 
the variable $n_i$, for $i=1,2,\ldots,N$,
as the height at site $i$ of a one-dimensional
interface. The condition $n_i\geq 0$ represents a ``hard wall.''
A general expression for the energy $\Eint\{n_i\}$
associated with an interface configuration $\{n_i\}$ is
\beq
\Eint = \sum_{i=1}^N H(n_i) + \sum_{i=1}^{N-1}V(n_i,n_{i+1}) 
\label{xEint}
\eeq
This expression was studied in reference \cite{vanLeeuwenHilhorst81}
for several different cases. The authors investigated 
in particular the linearly increasing on-site potential $H(n_i)=gn_i$
and the ``SOS'' interaction $V(n_i,n_{i+1})=2K|n_i-n_{i+1}|$,
in which $g$ is the ``gravitational constant'' and $2K$ the ``elastic''
energy. 
They found that for $g\to 0$ the interface width diverges as 
$\nav \sim g^{-1/3}$.

In the present work there is an effective interaction associated with
each pair of neighboring variables 
$n_i\equiv n$ and $n_{i+1}\equiv m$, 
mediated by the traced-out overlap $\ell_{i,i+1}$
and explicitly given by 
\beq
\Eint = \sum_i \log \tilT_{n_i,n_{i+1}}\,, 
\label{xEintspec}
\eeq
where we ignore boundary terms.
When (\ref{dTS}) and (\ref{xTnmb}) are substituted in (\ref{xEintspec})
and when we take $\nu\equiv\min(m,n)$ large, we find the expansion
\bea
\log\tilT_{nm} = \nu\log z\omega +\tfrac{1}{2}|n-m|\log 2z
                 + \mbox{cst} + \calO\big( (2/\omega)^\nu \big),
\quad \nu\to\infty,
\label{logTexp}
\eea
valid for $\omega>2$.
The coefficient $\tfrac{1}{2}\log 2z$ of the elastic term 
keeps a finite value when $z\to\zc=\omega^{-1}$. 
It multiplies $|n-m|$, and hence this term
is identical to the SOS interaction of
reference \cite{vanLeeuwenHilhorst81}.
The coefficient $\log z\omega$ of the potential term 
vanishes as $\sim(z-\zc)\omega$ for $z\to\zc$
and plays the role of the limit $g\to 0$ in reference  
\cite{vanLeeuwenHilhorst81}.
It multiplies, however, the minimum $\nu=\min(m,n)$ rather than the
symmetric sum $\tfrac{1}{2}(m+n)$. This difference, plus the fact that we have
expanded for large $\nu$, makes our model different from
that of reference \cite{vanLeeuwenHilhorst81}.
We may nevertheless speculate that with respect to their critical
behavior these two models are in the same universality class,
which explains the exponent $\alpha=1/3$ found in the simulation of
section \ref{sec_numericalTM}. 






\appendix

\end{document}